\documentclass{myaa}
\usepackage{graphics}
\begin{document}
\title{$1/f$ Noise and other Systematic Effects
in the Planck-LFI Radiometers}

\author{
    Michael Seiffert \inst{1}\and
    Aniello Mennella \inst{2}\and
    Carlo Burigana \inst{3}\and
    Nazzareno Mandolesi\inst{3}\and
    Marco Bersanelli \inst{4}\and
    Peter Meinhold\inst{5}\and
    Phil Lubin\inst{5}
} \offprints{A. Mennella}

\institute{
    Jet Propulsion Laboratory, California Institute of Technology,
Pasadena, CA 91109, USA\and
    IASF-CNR, Sezione di Milano, Via Bassini 15, 20133 Milan,
    Italy\and
    IASF-CNR, Sezione di Bologna, Via Gobetti 101, Bologna,
    Italy\and
    Universit\`a degli Studi di Milano, Via Celoria 16, 20133 Milan,
    Italy\and
    University of California at Santa Barbara, Physics Department, Santa Barbara, CA 93106,
    USA
    }

\abstract{We use an analytic approach to study the susceptibility
of the {\sc Planck} Low Frequency Instrument radiometers to
various systematic effects. We examine the effects of
fluctuations in amplifier gain, in amplifier noise temperature
and in the reference load temperature. We also study the effect
of imperfect gain modulation, non-ideal matching of radiometer
parameters, imperfect isolation in the two legs of the radiometer
and back-end $1/f$ noise. We find that with proper gain modulation
$1/f$ gain fluctuations are suppressed, leaving fluctuations in
amplifier noise temperature as the main source of $1/f$ noise. We
estimate that with a gain modulation factor within $\pm 1\%$ of
its ideal value the overall $1/f$ knee frequency will be
relatively small ($< 0.1 $Hz).}

\authorrunning{M. Seiffert et al.}

\maketitle

\section{\label{sec:introduction}Introduction}

{\sc Planck}\footnote{{\sc Planck} homepage:
http://astro.estec.esa.nl/Planck/} is a European Space Agency
(ESA) satellite mission to map spatial anisotropy and
polarization in the Cosmic Microwave Background (CMB) over a wide
range of frequencies with an unprecedented combination of
sensitivity, angular resolution, and sky coverage (Bersanelli et
al. \cite{bersanelli1}). Following the breakthrough of the COBE
discovery of CMB anisotropy (Smoot et al. \cite{smoot}, Bennett
et al. \cite{bennett}), and the MAP\footnote{MAP homepage:
http://map.gsfc.nasa.gov} satellite launched in June 2001, {\sc
Planck} will be the third generation space mission dedicated to
CMB observations.

The data gathered by these missions will revolutionise modern
cosmology by a precise determination of the fundamental
cosmological parameters which govern the present expansion rate
of the universe, the average density of the universe, the amount
of dark matter, and the nature of the seed fluctuations from
which all structures in the universe arose (see, e.g., Hu et al.
\cite{Hu}, Scott et al. \cite{scott}, White et al. \cite{white}
for recent reviews on CMB anisotropy) and will provide at the
same time full sky surveys at essentially unexplored frequencies,
with fundamental implications for a large area of problems in
astrophysics (De Zotti et al. \cite{dezotti}).

{\sc Planck} consists of a High Frequency Instrument (HFI) (Puget
et al. \cite{puget}) and a Low Frequency Instrument (LFI)
(Mandolesi et al. \cite{mandolesi}) observing the sky through a
common telescope. While the MAP radiometers measure temperature
differences between two widely separated regions of the sky
through a pair of symmetric back-to-back telescopes, the {\sc
Planck}~LFI radiometers are designed to measure differences
between the sky signal and a stable internal cryogenic reference
load. The LFI scheme takes advantage of the presence in the focal
plane of the HFI front end unit, which is cooled to approximately
4 K as an intermediate cryo stage for the 0.1 K bolometer
detectors.

The LFI radiometer design is a modified correlation receiver
(Blum \cite{blum}, Colvin \cite{colvin}, Bersanelli et al.
\cite{bersanelli2}), realised with High Electron Mobility
Transistor (HEMT) amplifiers at 30, 44, 70 and 100 GHz. The
modification is that the temperature of the reference load can be
made significantly different from the sky temperature. To
compensate for the offset (a few K in nominal conditions), a {\sl
gain modulation factor}, $r$, is used to null the output signal
in order to minimise sensitivity to RF gain fluctuations and
achieve the lowest white and $1/f$ noise in the output.

Obtaining data streams with low $1/f$ noise is of primary importance
in order to achieve the LFI scientific objectives. In fact
excessive $1/f$ noise would degrade the quality of the measured
data (Janssen et al. \cite{janssen}) 
by increasing the effective rms noise and the uncertainty in
the power spectrum at low multipole values. Such effects can be avoided if the post
detection knee frequency $f_k$ (i.e. the frequency at which the
$1/f$ contribution and the ideal white noise contribution are
equal) is significantly lower than the spacecraft rotation
frequency ($f_{\rm spin}\sim 0.017$ Hz). For values of $f_k$
greater than $f_{\rm spin}$ it is possible to mitigate such
effects by applying appropriate {\em destriping} and {\em map
making} algorithms\footnote{see Burigana et al. \cite{burigana1},
Delabrouille \cite{delabrouille}, Maino et al. \cite{maino2}, for
details about {\em destriping} and Dor\'e et al. \cite{dore},
Natoli et al. \cite{natoli}, for details about {\em map-making}
algorithms.} to the time ordered data (Maino et al. \cite{maino1}).

If the knee frequency is sufficiently low (i.e.
$f_k\leq 0.1$~Hz), with the application of such algorithms it is possible to
maintain the increase in rms noise within few \% of the white
noise, and the power increase at low multipole values (i.e.
$l\leq 200$) at a very low level ($\sim$ two order of magnitude less
than the CMB power). If, on the other hand, the knee frequency is high
(i.e. $\gg 0.1$~Hz) then even after destriping the degradation of 
the final sensitivity is of several tenths of \% and the excess 
power at low multipole values is significant 
(up to the same order of the CMB power for 
$f_k\sim 10$~Hz, Bersanelli et al., \cite{bersanelli2002}).
Therefore, careful attention to instrument design, 
analysis, and test is essential in order to achieve a low $1/f$ 
noise knee frequency.

In this paper, we analyse the most important systematic effects
due to non-ideal behaviour of components in the LFI radiometer
signal chain, and estimate their impact on the post detection
knee frequency. In section~\ref{sec:analytic_model_LFI} we
present a general analytical description of the {\sc Planck}-LFI
radiometers to derive formulas for the radiometer power output
and sensitivity in the two cases of perfectly balanced and
slightly unbalanced radiometer. Here we also show that under
quite general assumptions, the radiometer sensitivity does not
depend on the reference load temperature. In
section~\ref{sec:systematic_effects} we analyse the impact of
various systematic effects on the post-detection knee frequency
showing that with proper gain modulation it is possible to keep
the radiometer $1/f$ noise to a very low level also in the
presence of different non-ideal behaviours (e.g., gain and noise
temperature imbalance, imperfect gain modulation). For sake of
conciseness, we transfer part of the formalism to the appendices.
Finally, in section~\ref{sec:conclusions} we summarise our
results and discuss briefly their implications for Planck
observations.

\section{\label{sec:analytic_model_LFI}Analytic model of LFI pseudo-correlation radiometers\label{sec:analytic_model}}
\subsection{\label{subsec:radiometer_architecture}Radiometer architecture\label{subsec:architecture}}
\begin{figure}[here]
\begin{center}
\resizebox{9. cm}{!}{\includegraphics{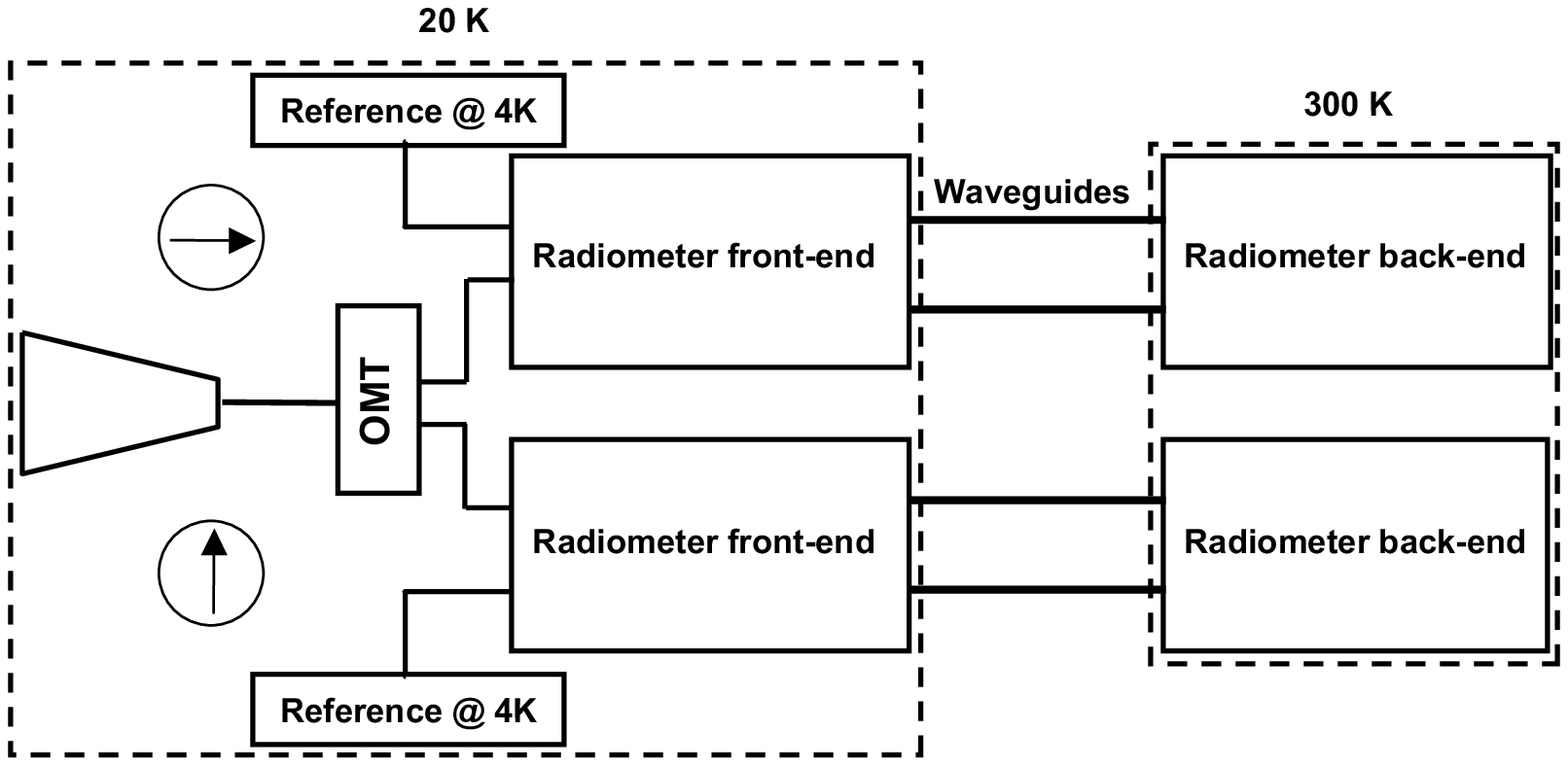}}
\end{center}
\caption{Schematic of the Planck LFI radiometer design. Each feed
horn is connected to two radiometers constituted of a 20~K
front-end and a warm back end at 300~K. In our design the
reference load has a temperature of 4~K.\label{fig:LFI_schematic}}
\end{figure}

In Figure \ref{fig:LFI_schematic} we show a schematic of the
baseline LFI radiometer design. In this design, each feed-horn is
connected to a Radiometer Chain Assembly consisting of an
actively-cooled 20~K front-end connected to a 300~K back-end via
waveguides.

\begin{figure}[here]
\begin{center}
\resizebox{9. cm}{!}{\includegraphics{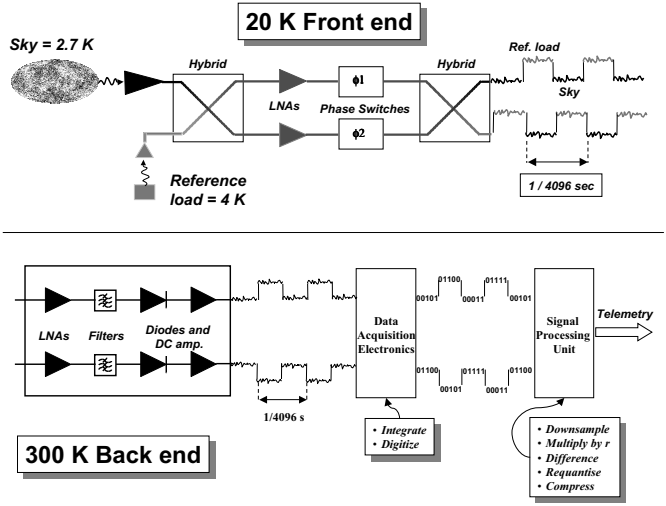}}
\end{center}
\caption{Details of the front-end (upper figure) and back-end
(lower figure) of LFI radiometers. \label{fig:front_back_end}}
\end{figure}

In the front-end part (see top part of
Fig.~\ref{fig:front_back_end}) the radiation entering the
feed-horn is separated by an OrthoMode Transducer (OMT) into two
perpendicular linearly polarised components that propagate
independently through two parallel radiometers. In each
radiometer, the sky signal and the signal from a stable reference
load at $\sim$4~K are coupled to cryogenic low-noise High
Electron Mobility Transistor (HEMT) amplifiers via a $180^\circ$
hybrid. One of the two signals then runs through a switch that
applies a phase shift which oscillates between 0 and $\pi$ at a
frequency of 4096~Hz. A second phase switch will be present for
symmetry on the second radiometer leg; this switch will not
introduce any phase shift in the propagating signal. Therefore 
it will not be considered in our analysis. The signals are then
recombined by a second 180$^\circ$ hybrid coupler, producing an
output which is a sequence of signals alternating at twice the
phase switch frequency.

In the back-end of each radiometer (see bottom part of
Fig.~\ref{fig:front_back_end}) the RF signals are further
amplified, filtered by a low-pass filter and then detected. After
detection the sky and reference load signals are integrated,
digitised and then differenced after multiplication of the
reference load signal by a so-called {\em gain modulation
factor}, $r$, which has the function to make the sky-load
difference as close as possible to zero.

According to this architecture each radiometer will produce two
independent streams of sky-load differences; the final
measurement is provided by a further averaging of these
differenced data samples between the two radiometer legs.

The LFI pseudo-correlation design offers two main advantages: the
first is that the radiometer sensitivity does not depend (to
first order) on the level of the reference signal; the second is
provided by the fast switching that reduces the impact of $1/f$
fluctuations of back-end amplifiers. In the next subsection we
present the analytical description of this radiometer design. In
particular we derive formulas for the power output and
sensitivity in the two approximations of (i) perfectly balanced
and (ii) slightly unbalanced radiometer. Then we show how the
radiometer sensitivity is not dependent, to first order, on the
reference load temperature and we also provide an estimate of this
dependence in the case of slightly unbalanced radiometer.

In order to improve readability we have kept the mathematical
treatment as simple as possible, relegating the full definition
of long formulas to appendices.

\subsection{Analytic model of LFI radiometers\label{subsec:analytic_model}}

\subsubsection{\label{subsubsec:input_output}Input-output signal sequence}

Referring to the schematic in Fig. \ref{fig:front_back_end}, we
can define the following transfer functions for the different
radiometer components.

\begin{eqnarray}
&&f_{\rm hyb}:
    \{x,y\}\rightarrow
    \left\{\frac{x+y}{\sqrt{2}},\frac{x-y}{\sqrt{2}}\right\} \nonumber \\
&&\mbox{}\nonumber\\
&&f_{\rm amp}^{\rm FE}:
    \{x,y\}\rightarrow\nonumber\\
&&\,\,\,\,\,\,\,\,\,\,\,\,\,\,\,\,\,\,\,\,\,\,\,\,\,\,\,\,\,\,\,\,\,\,\,\,
    \left\{g_{F_1}(x+n_{F_1})e^{i\phi_{F_1}}, g_{F_2}(y+n_{F_2})
    e^{i\phi_{F_2}}\right\}\nonumber\\
&&\mbox{}\nonumber\\
&&f_{\rm sw}:
    \{x,y\} \rightarrow
                \left\{x,\sqrt{A_j} e^{i\theta_j}y\right\} \,\,\,\,\,{\mbox with}\,j=1,2\\
&&\mbox{}\nonumber\\
&&f_{\rm amp}^{\rm BE}:
    \{x,y\}\rightarrow\nonumber\\
&&\,\,\,\,\,\,\,\,\,\,\,\,\,\,\,\,\,\,\,\,\,\,\,\,\,\,\,\,\,\,\,\,\,\,\,\,
    \left\{g_{B_1}(x+n_{B_1})e^{i\phi_{B_1}},
    g_{B_2}(y+n_{B_2})e^{i\phi_{B_2}}\right\}\nonumber
\label{eq:transfer_functions}
\end{eqnarray}

 \noindent where:

\begin{itemize}
\item $x(t)$ and $y(t)$ represent the voltages at the sky
and reference horns respectively;

\item $g_{F_1}$, $g_{F_2}$, $g_{B_1}$ and $g_{B_2}$ are the
voltage gains of front-end and back-end amplifiers;

\item $n_{F_1}(t)$, $n_{F_2}(t)$, $n_{B_1}(t)$ and
$n_{B_2}(t)$ represent the noise voltages of the front-end and
back-end amplifiers;

\item $\phi_{F_1}$, $\phi_{F_2}$, $\phi_{B_1}$ and
$\phi_{B_2}$ represent the signal phases after the amplifiers;

\item $\theta_1$ and $\theta_2$ represent the phase shifts
in the two switch states (in LFI baseline $\theta_1 = 0$ and
$\theta_2 = \pi$);

\item $A_1$ and $A_2$ represent the fraction of the signal amplitudes 
that is transmitted after
the phase switch in the two switch states. For lossless switches
$A_{1,2}=1$.\end{itemize}

The output signal at the two radiometer legs is given by:

\begin{eqnarray}
s_{j,1}&=& e^{i \phi_{B_1}}g_{B_1}\left\{
n_{B_1}+\frac{1}{\sqrt{2}} \left[ e^{i
\phi_{F_1}}g_{F_1}\left(n_{F_1}+\frac{x+y}{\sqrt{2}}\right)+
\right.\right.\nonumber\\
&+&\left.\left.\sqrt{A_j}e^{i(\theta_j+\phi_{F_2})}g_{F_2}\left(n_{F_2}+
\frac{x-y}{\sqrt{2}}\right)\right]\right\} \nonumber \\
&&\\
s_{j,2}&=& e^{i \phi_{B_2}}g_{B_2}\left\{
n_{B_2}+\frac{1}{\sqrt{2}} \left[ e^{i
\phi_{F_1}}g_{F_1}\left(n_{F_1}+\frac{x+y}{\sqrt{2}}\right)+
\right.\right.\nonumber\\
&-&\left.\left.\sqrt{A_j}e^{i(\theta_j+\phi_{F_2})}g_{F_2}\left(n_{F_2}+
\frac{x-y}{\sqrt{2}}\right)\right]\right\}\nonumber
\label{eq:xout}
\end{eqnarray}

\noindent where the subscript $j=1,2$ indicates the two phase
switch states.

\subsubsection{Radiometer power output and sensitivity}

Our first step is to write an expression for the radiometer power
output at a given time $t$, considering that in our scheme we take
the difference between the sky and load signals. If we indicate
with $p_{\, l}(t)$ the power output at each of the two radiometer
legs (where $l=1,2$) and with
$p\,(t)$ the radiometer power output we have that:

\begin{eqnarray}
\label{eq:_power_output_general} &&
p_{\,1}(t)=a\left(|s_{1,1}|^2(t)-r|s_{2,1}|^2(t+\Delta t_{\rm
sw})\right)\nonumber\\
&& p_{\, 2}(t)=a\left(|s_{2,2}|^2(t+\Delta t_{\rm
sw})-r|s_{1,2}|^2(t)\right)\\
&& p\,(t)=1/2\left[p_1(t)+p_2(t)\right]\nonumber
\end{eqnarray}

\noindent where $a$ is the constant of proportionality of the
square-law detectors, $\Delta t_{\rm sw}\sim 122$~$\mu$s
corresponds to half of the phase switch period and $r$ is the
{\em gain modulation factor} which in LFI scheme is applied in
software. The expanded form of $p_{\, 1}(t)$ is reported in
equation (\ref{eq:p(t)}); a similar relationship holds for
$p_2(t)$. The assumption of identical values of $a$ in the above
equation can be made without loss of generality; any difference in
actual values between the two legs could be absorbed into
back-end gain differences for the purposes of this analysis.

In our treatment we will show how the gain modulation factor (a
real number generally $\leq 1$) can be tuned properly in order to
have an approximately null power output, which leads to almost
complete suppression of the dependency of the radiometer
sensitivity on the reference load signal, and minimal impact of
gain fluctuations in the front-end amplifiers. Note that in
equation (\ref{eq:_power_output_general}) the sky and load
signals appearing in the difference are sampled at slightly
different times, which are relative to the two different phase
switch states; this is relevant when we consider the effect of
1/$f$ fluctuations of the back-end amplifiers. The effect of
these instabilities can be practically eliminated by using a
phase switch frequency $f_{\rm sw}$ much greater than the 1/$f$
noise knee frequency; in section \ref{sec:sensitivity_BE_amps} we
will show that the baseline phase switch frequency of $\sim
4$~KHz considered for LFI leads to a very low level of back-end
1/$f$ noise.

Let us now calculate the time average of $p_{\, l}(t)$ integrated
over the bandwidth, $\beta$, i.e. $\overline p_{\,
l}=\int_{\beta}\int_{\Delta t_{\rm sw}}p_{\, l}(t)dt d\nu$.
Considering that signal and amplifier noises are uncorrelated we
have that cross-correlation terms vanish; therefore we can write
$\overline p_1$ and $\overline p_2$ in the following compact form:

\begin{eqnarray}
\label{eq:p_av}
\overline p_1 &&= a k \beta G_{B_1}\times\nonumber\\
&&\times\left[
    T_x(\hat{G}-r\hat{I}) - r
    T_y\left(\tilde{G}-\frac{1}{r}\tilde{I}\right)+\hat{T}_{n_1}-r
    \tilde{T}_{n_1} \right]\nonumber\\
&&\mbox{}\\
\overline p_2 &&= a k \beta G_{B_2}\times\nonumber\\
&&\times\left[
    T_x(\tilde{G}-r\tilde{I}) - r
    T_y\left(\hat{G}-\frac{1}{r}\hat{I}\right)+\tilde{T}_{n_2}-r
    \hat{T}_{n_2} \right]\nonumber
\end{eqnarray}

\noindent where:

\begin{eqnarray}
&&\hat{G}=\frac{1}{4}\left[ G_{F_1} + A_1 G_{F_2} +2 \sqrt{A_1
G_{F_1} G_{F_2}} \cos(\theta_1+\phi)\right]\nonumber\\
&&\hat{I}=\frac{1}{4}\left[G_{F_1} + A_2 G_{F_2} +2 \sqrt{A_2
G_{F_1} G_{F_2}} \cos(\theta_2+\phi)\right]\nonumber\\
&&\tilde{G}=\frac{1}{4}\left[ G_{F_1} + A_2 G_{F_2} -2 \sqrt{A_2
G_{F_1} G_{F_2}} \cos(\theta_2+\phi)\right]\nonumber\\
&&\tilde{I}=\frac{1}{4}\left[ G_{F_1} + A_1 G_{F_2} -2 \sqrt{A_1
G_{F_1} G_{F_2}} \cos(\theta_1+\phi)\right]\nonumber\\
&&\hat{T}_{n_l}=\frac{1}{2}(G_{F_1} T_{n_{F_1}} + A_1 G_{F_2}
T_{n_{F_2}} + 2\,T_{n_{B_l}})\nonumber\\
&&\tilde{T}_{n_l}=\frac{1}{2}(G_{F_1} T_{n_{F_1}} + A_2 G_{F_2}
T_{n_{F_2}} + 2\, T_{n_{B_l}})\nonumber
\end{eqnarray}

In equation (\ref{eq:p_av}) the terms $T_x$, $T_y$, $T_{n_{F_1}}$,
$T_{n_{F_2}}$, $T_{n_{B_2}}$ and $T_{n_{B_2}}$ represent the sky,
load and amplifier noise temperatures which are defined by
relationships like $k \beta T_x = x^2$($k$ is the Boltzmann
constant), $G_{F_1,F_2,B_1,B_2}=g_{F_1,F_2,B_1,B_2}^2$ are the
amplifier power gains, $\beta$ is the bandwidth and $\phi =
\phi_{F_2}-\phi_{F_1}$ is the phase mismatch of the front-end
amplifiers.

In the case in which radiometer parameters (gains, noise
temperatures, etc.) depend on the frequency, equation
(\ref{eq:p_av}) is still valid provided that we use values
averaged over the bandwidth, i.e. for each parameter $P$ on the
left-hand side of the above definitions: $P\rightarrow
\overline P = \frac{1}{\nu}\int_{\nu_0 -\frac{\beta}{2}}^{\nu_0
+\frac{\beta}{2}}P(\nu)d\nu$.

We now see that in order to null the output at each radiometer
branch output (i.e. make $\overline{p_l} = 0$), we must adjust $r$
to a proper value, denoted $r^*$. In the general case discussed
above we have that:

\begin{eqnarray}
&&\overline p_1=0\,\Longrightarrow \, r=r^*_1=\frac{T_x \hat{G} +
T_y \tilde{I} + \hat{T}_{n_1}}{T_y \tilde{G} + T_x \hat{I} +
\tilde{T}_{n_1}}\nonumber\\
&&\overline p_2=0\,\Longrightarrow \, r=r^*_2=\frac{T_x \tilde{G}
+ T_y \hat{I} + \tilde{T}_{n_2}}{T_y \hat{G} + T_x \tilde{I}
+\hat{T}_{n_2}} \label{eq:rstar}
\end{eqnarray}

It should be noted that both the sky temperature and the
instrumental parameters are expected to undergo changes of order
of few mK. Thermal variations and aging are likely to drive
instrumental changes which are expected to occur on time scales
of days to months. The main source of change in $T_x$ is the CMB
dipole signature, which introduces a spin-synchronous (1~rpm)
temperature modulation $\delta T_x \leq 3.5$~mK. In principle it
could be possible to implement feedback schemes in order to vary
$r$ almost in real time and have a constant zero output\footnote{In 
this case the astrophysical information would be
recovered by recording the gain change needed to maintain the radiometer
balance.}; however, this would complicate the scheme and
possibly introduce additional sources of systematics. We
anticipate that adjusting $r$ on the timescale of a few days will
be sufficient, so that we can consider $r$ constant in the following
calculations. This issue is discussed in more detail in a forthcoming
paper currently in preparation.

The radiometer sensitivity $\Delta T$ (i.e. the minimum signal
that can be detected in a bandwidth $\beta$ and integration time
$\tau$) can be calculated following the approach outlined in
appendix~\ref{app:sensitivity} where we also report the complete
algebraic form.

\subsubsection{Approximations\label{sec:approximations}}

In this section we derive some approximations of the general
equations presented in appendices \ref{app:power_output} and
\ref{app:sensitivity}.

\paragraph{Perfectly balanced radiometer.\label{par:ideal_case}}

The first, zero-order approximation is relative to the ideal case
of a perfectly balanced radiometer. We assume that the radiometer
components are ideally matched, but that there still is a
temperature offset between the sky and reference load, and that
there are still noise and gain fluctuations present in the
amplifiers. This approximation can be derived by setting $G_{F_1}
= G_{F_2} \equiv G_{\rm F}$, $G_{B_1} = G_{B_2} \equiv G_{\rm
B}$, $T_{n_{F_1}}=T_{n_{F_2}}\equiv T_{n_{\rm F}}$,
$T_{n_{B_1}}=T_{n_{B_2}}\equiv T_{n_{\rm B}}$, $\theta_1 = 0$,
$\theta_2 = \pi$, $\phi=0$, $A_1=A_2=1$. In this case the
radiometer power output and sensitivity can be written in the
following simple form:

\begin{eqnarray}
&& \overline p \rightarrow \overline p_0 =a k \beta G_{\rm
F}G_{\rm B} \left[ T_x + T_{n_{tot}} -r (T_y + T_{n_{tot}})
\right]
\nonumber \\
&& r^*_1, r^*_2 \rightarrow r^*_0=\frac{T_x+T_{n_{tot}}}{T_y+T_{n_{tot}}}\nonumber\\
&& \Delta T|_{r\equiv r^*}\rightarrow \Delta T_0|_{r\equiv
r_0^*}=\nonumber\\
&&=\sqrt{\frac{2}{\beta
\tau_0}}(T_x+T_{n_{tot}})\sqrt{1-\frac{(T_{n_{\rm B}}/G_{\rm
F})^2}{(T_x+T_{n_{tot}})(T_y+T_{n_{tot}})}} \label{eq:basic_case}
\end{eqnarray}

\noindent where $T_{n_{tot}}=T_{n_{\rm F}}+T_{n_{\rm B}}/G_{\rm
F}$ and $\tau_0$ represents the integration time needed to obtain
a sky-load measurement. Note that the above expressions are
relative to power output and sensitivity of the complete
radiometer, i.e. after averaging the two radiometer legs; the
main advantage of this averaging is that any first-order
dependency of the knee frequency on the mismatch in signal amplitudes
after the phase switch is cancelled.

From equation (\ref{eq:basic_case}) we see that the radiometer
sensitivity is not fully independent of the reference load
temperature, $T_y$, because the signals in the two radiometer
branches are correlated upstream of the back-end amplification.
On the other hand for LFI typical parameters, the term
$\frac{(T_{n_{\rm B}}/G_{\rm
F})^2}{(T_x+T_{n_{tot}})(T_y+T_{n_{tot}})}$ is of the order
$10^{-4}$ $-$ $10^{-5}$. Therefore the sensitivity $\Delta
T_0|_{r\equiv r^*_0}$ in equation (\ref{eq:basic_case}) can be
approximated by:

\begin{equation}
\Delta T_0|_{r\equiv r^*_0}\approx \sqrt{\frac{2}{\beta
\tau}}(T_x+T_{n_{tot}})\label{eq:basic_case_approx}
\end{equation}

\noindent which is independent from $T_y$. In the following of
this paper we will always refer to the radiometer sensitivity
calculated for $r = r^*_0$ even if not explicitly indicated.

\paragraph{Slightly unbalanced
radiometer.\label{par:slightly_unbalanced}} We will now consider
the case of a slight imbalance in the radiometer front-end,
assuming $G_{B_1}=G_{B_2}=1$ and $T_{n_{B_1}}=T_{n_{B_2}}=0$. The
impact of 1/$f$ noise from the back-end amplifiers will be
analysed in section~\ref{sec:sensitivity_BE_amps}.

The first-order approximation of a {\em slightly unbalanced
front-end} can be described mathematically by setting:

\begin{equation}
\begin{array}{ll}
G_{F_1} = G & G_{F_2} = G(1+\epsilon_G)\\
T_{n_{F_1}} = T_n & T_{n_{F_2}}=T_n(1+\epsilon_{T_n})\\
A_1 = 1-\epsilon_{A_1} & A_2=1-\epsilon_{A_2}\\
\phi = \epsilon_{\phi} \\
\theta_1 = \epsilon_{\theta_1} & \theta_2 =
\pi(1+\epsilon_{\theta_2}) \label{eq:slight_unbalance}
\end{array}
\end{equation}

In equation (\ref{eq:slight_unbalance}) we assume that
$\epsilon_G < 0.3$ (i.e. we assume a gain mismatch $\leq 1$~dB)
and that all other $\epsilon$ parameters are $< 0.1$. Then we
expand equations (\ref{eq:p_av}) and (\ref{eq:sensitivity}) in
series at the second order in $\epsilon_G$ and at the first order
in the other parameters, obtaining the following approximate
equations for the average power output, $\overline p$, and the
sensitivity, $\Delta T$ (the parameters $\alpha_1$ through
$\alpha_4$ are defined in
appendix~\ref{app:definition_alpha_pars}):

\begin{eqnarray}
\label{eq:1st_order_radiometer} \overline p &\approx& \overline
p_0 \left[ 1 + 1/2\left(\epsilon_G -
\alpha_1\epsilon_G^2-(\epsilon_{A_1}+\epsilon_{A_2})/2+\alpha_2\,\epsilon_{T_n}\right)\right]
\nonumber\\
&&\mbox{}\nonumber\\
\Delta T &\approx&
    \Delta T_0\left(
        1+1/4 \, \alpha_3\,\epsilon_{T_n}+1/16 \,\alpha_4\,
        \epsilon_G^2
    \right)\approx\\
&\approx&\Delta T_0\left(
        1+1/4 \, \alpha_3\,\epsilon_{T_n}
    \right)\nonumber
\end{eqnarray}

\noindent where we have neglected the term $1/16
\,\alpha_4\,\epsilon_G^2$ (of the order of
$\simeq\,10^{-3}\epsilon_G^2$) with respect to
$1/4\,\alpha_3\,\epsilon_{T_n}$ (of the order of $\simeq
0.4\,\epsilon_{T_n}$).

From equation (\ref{eq:1st_order_radiometer}) it is apparent that
the main additional contribution to the ideal sensitivity comes
from the unbalance in the front-end noise temperature. For LFI
typical values the sensitivity is degraded by a factor in the
range $(0.4-0.5)\times\epsilon_{T_n}$ which means, in other
words, that with a noise temperature match better than 5\% it is
possible to maintain the sensitivity degradation at levels below
1\%. A second consideration is that small non-idealities in the
radiometer chains introduce only a very weak dependence of the
sensitivity on the reference load temperature; in fact from
equation (\ref{eq:1st_order_radiometer}) it follows that
$\left|\frac{\partial \Delta T}{\partial
T_y}\right|=\epsilon_{T_n}\sqrt{\frac{1}{2\beta\tau}}
\frac{T_n(T_x+T_n)}{2(T_y+T_n)^2}\approx 1.5\times
10^{-5}\epsilon_{T_n}$.

\section{Susceptibility to various systematic effects\label{sec:systematic_effects}}

In this section we study the sensitivity to some potential
systematic effects arising from fluctuations in the radiometers;
in fact fluctuations in any of the terms appearing in
equation~(\ref{eq:p_av}) will lead to a change in the observed
signal which can mimic a ``true" sky fluctuation. If we denote
with $\Delta T_{\rm eq}$ the spurious signal fluctuation induced
by a variation in a generic radiometer parameter $w$ we have that:

\begin{equation}
{\partial \bar p \over \partial T_x} \Delta T_{\rm eq} =
{\partial \bar p \over \partial w} \Delta w
\label{eq:general_approach}
\end{equation}

In the following sections of this paper, we calculate the
magnitude of these effects under the assumption that the
fluctuations in the various parameters are uncorrelated. Before
calculating these effects we briefly examine the expected
magnitude of gain and noise temperature fluctuations.

\subsection{Intrinsic HEMT amplifier noise characteristics}

Cryogenic HEMT amplifiers are well known to have $1/f$ type gain
fluctuations, and from this we can infer that they also have
$1/f$ type fluctuations in noise temperature (Pospieszalski
\cite{pospieszalski}, Wollack \cite{wollack}, Jarosik
\cite{jarosik}). The level of these fluctuations can vary
considerably among amplifiers and depends on the details of
device fabrication, device size, circuit design, and other
factors.  Because of this, we will adopt an empirical model for
the fluctuations. We can write the $1/f$ spectrum of the gain
fluctuations as:
\begin{equation}
\frac{\Delta G(f)}{G} = \frac{C}{\sqrt{f}} \label{eq:DG/G}
\end{equation}
where $C$ represents a constant normalization factor. Similarly,
we can write the noise temperature fluctuations as
\begin{equation}
\frac{\Delta T_n(f)}{T_n}  = \frac{A}{\sqrt{f}} \label{eq:deltat}
\end{equation}
where $A$ is the normalization constant for noise temperature
fluctuations. Here $\Delta T(f) / T$ has units of Hz$^{-1/2}$ and
$A$ is dimensionless.  From elementary statistical
considerations, we can infer that $A=C/2{\sqrt{N_s}}$, where
$N_s$ is the number of stages in the amplifier.

A normalization of $A \simeq 1.8 \times 10^{-5}$ (see above
references) is appropriate for the 30 and 44 GHz radiometers. For
the 70 and 100 GHz radiometers it will be necessary to use HEMT
devices with a smaller gate width to achieve the lowest amplifier
noise figure. We expect that the gate widths will be roughly
$1/2$ that of the devices used for the lower frequency
radiometers and this will lead to fluctuations that are roughly a
factor of $\sqrt{2}$ higher. For the higher frequencies we will
therefore adopt a normalization of $A=2.5 \times 10^{-5}$. Note
that values for $A$ given here should be regarded as estimates
rather than precise values, because in general $A$ will be
different for any particular device and will generally depend on
the physical temperature of the amplifier.

While the calculations in this paper assume the above simple
functional form for $\Delta T(f)$, it is straightforward to
repeat the calculations with a more detailed spectral shape.

\subsection{Front-end amplifier fluctuations\label{sec:sensitivity_FE_amps}}

\subsubsection{Sensitivity to noise temperature fluctuations\label{sec:sensitivity_FE_amps_Tn}}

In this section, we will calculate the change in the output
signal for a small change in the noise temperature of the
front-end amplifiers. Using equation (\ref{eq:general_approach})
we have that the change in input signal induced by a fluctuation
in $T_{n_{F_1}}$ is given by:

\begin{equation}
\Delta T_{\rm eq}(f) =\Delta T_n(f) \frac{\partial \overline p /
\partial T_{n_{F_1}}}{\partial\overline p /
\partial T_x} \nonumber
\label{eq:effect_Tn_1}
\end{equation}

Because both amplifiers (which have uncorrelated noise) can
contribute to the change in input signal, we have that the change
in input signal induced by noise temperature fluctuations in both
amplifiers is:

\begin{eqnarray}
\Delta T_{\rm eq}(f) &=& \Delta T_n(f) \left[ \left(\frac{\partial
\overline p / \partial T_{n_{F_1}}}{\partial\overline p /
\partial T_x}\right)^2+\left(\frac{\partial \overline p / \partial
T_{n_{F_2}}}{\partial\overline p /
\partial T_x}\right)^2\right]^\frac{1}{2}\nonumber\\
&\approx& \Delta T_n(f)\,\frac{r - 1}{\sqrt{2}}\,\left(1 + \frac{3
+ r}{16}\, {{{\epsilon }_G^2}}\right) \label{eq:effect_Tn1_Tn2}
\end{eqnarray}

If the reference load is at the same temperature as the sky, then
$r=1$ and the effect vanishes.

Now we calculate the post-detection frequency, $f_k$, at which
the contributions from noise temperature fluctuations are equal
to the white noise from an ideal radiometer:

\begin{equation}
\Delta T_{\rm eq}(f_k)|_{r=r^*_0} = \Delta
T(f)\label{eq:calc_fk_Tn}
\end{equation}

\noindent From equation (\ref{eq:1st_order_radiometer}) we can
write the sensitivity spectral density as:

\begin{equation}
\Delta T(f) = \Delta T_0(f)\left(
        1+1/4 \, \alpha_3\,\epsilon_{T_n}
    \right)
\label{eq:sens_spectrum}
\end{equation}

\begin{figure*}[htpb]
\begin{center}
\resizebox{13.5 cm}{!}{\includegraphics{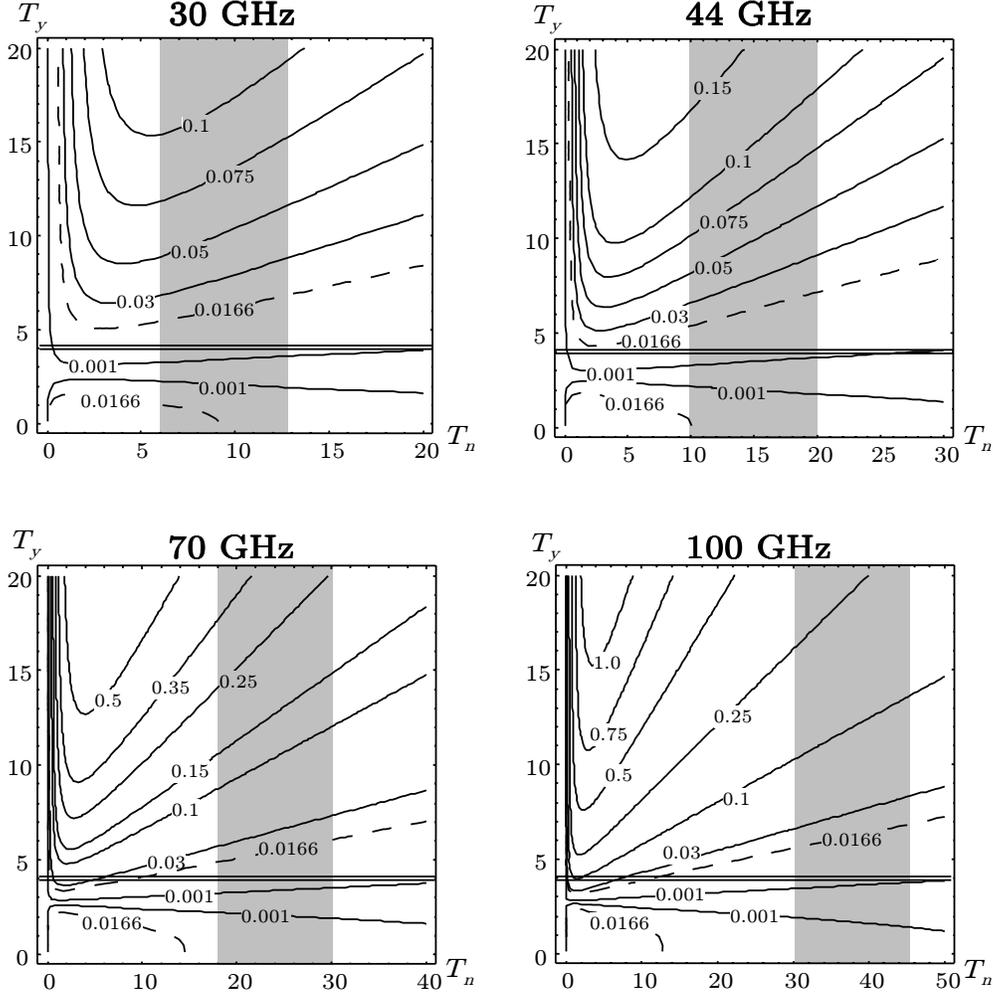}}
\end{center}
\caption{curves of equal $f_k$ (in Hz) on the plane $T_y$ (K,
thermodynamic temperature), $T_n$ assuming a thermodynamic sky
temperature of 2.7~K. Each panel refers to a different frequency
channel. The dashed contour refers to values for which the knee
frequency is equal to the spin frequency ($f_{spin}=0.0166$~Hz).
The graphs also show the range of typical LFI noise temperature
values (grey area) and the nominal reference load temperature
(4~K $-$ double horizontal line).}\label{fig:fk_Tn}
\end{figure*}

\noindent where $\Delta T_0(f) = \sqrt{\frac{2}{\beta}}(T_x+T_n)$.
Now with simple algebra we can rewrite equation
(\ref{eq:calc_fk_Tn}) in terms of $\frac{\Delta T_n}{T_n}$ which
equals $A/\sqrt{f}$ (see equation (\ref{eq:deltat})). A further
expansion in $\epsilon$ yields:

\begin{equation}
f_k(T_n) = f_k^0(T_n)\left(1-1/2\,\alpha_3 \,
\epsilon_{T_n}+1/8\,(r+3)\,\epsilon_G^2\right) \label{eq:kneefreq}
\end{equation}

\noindent where $f_k^0(T_n)=
   \beta \,
   {\left[ \frac{A\,\left( 1 - r \right) \,{T_n}}
       {2\,\left( {T_x} + {T_n} \right) } \right] }^2$
and $\alpha_3$ is defined in appendix
\ref{app:definition_alpha_pars}.

Let us now consider first the ideal case, in which
$f_k(T_n)=f_k^0(T_n)$. Assuming a 20\% bandwidth and a
thermodynamic sky temperature of 2.7 K, we can calculate knee
frequencies for several choices of $T_y$ and $T_n$. Results are
summarised graphically in Figure \ref{fig:fk_Tn}, where we show
contour plots of constant $f_k$ as a function of $T_y$ (given in
thermodynamic temperature) and $T_n$. The contour relative to
$f_k=f_{\rm spin}=0.0166$~Hz is shown with a dashed line. The
graphs also show (for each frequency) the range of typical noise
temperature values (grey area) and the nominal reference load
temperature (4~K $-$ double horizontal line). These results show
that at all frequencies the expected knee frequency arising from
noise temperature fluctuations is $f_k^0(T_n)\sim$ 3 $-$ 5~mHz,
i.e. about 4 to 7 times less than the spin frequency.

In the limits of our approximations (i.e. a noise temperature
match better than 10\% and a gain match better than 1~dB)
radiometer non idealities determine a correction to the
zero-order knee frequency that is within $\pm 10\%$.

\subsubsection{Sensitivity to gain fluctuations
\label{sec:sensitivity_FE_amps_G}}

In this section we calculate the radiometer sensitivity to
front-end gain fluctuations and show that with proper gain
modulation the effect is negligible compared to the effect
induced by fluctuations in $T_n$.

Proceeding similarly to the previous section the post-detection
knee frequency of $1/f$ noise caused by gain fluctuations in the
front-end amplifiers (in the case of a perfectly balanced
radiometer) is:

\begin{equation}
f_k^0(G)=\frac{1}{2}\,C^2
\beta\frac{\left[T_x+T_n-r(T_y+T_n)\right]^2}{(T_x+T_n)^2+r^2\,(T_y+T_n)^2}
 \label{eq:kneefreq_G}
\end{equation}

\noindent which shows that with proper gain modulation
($r=r^*_0$) the radiometer is insensitive to gain fluctuations,
i.e. $f_k^0(G)=0$. If we calculate $f_k(G)$ in the general case
with the usual series expansion we have:

\begin{equation}
\left. f_k(G)\right|_{r=r^*_0}\approx \, \beta {\left[ \frac{C
\left( {T_x} - {T_y} \right) \left(
                2{T_n} + {T_x} + {T_y} \right) }{8\left( {T_n} + {T_x} \
\right) \left( {T_n} + {T_y} \right) } \right]
}^2 \, \epsilon_G^2\label{eq:kneefreq_G_r!=r*}
\end{equation}

From equation (\ref{eq:kneefreq_G_r!=r*}) we see that for typical
LFI parameters and $\epsilon_G\sim\pm\,0.15$ (which corresponds
to a gain mismatch of the order of $\sim\pm\, 0.5$~dB) the knee
frequency  $f_k(G)|_{r=r^*} < 1$~mHz, which indicates that with
proper gain modulation the $1/f$ noise is still dominated by noise
temperature fluctuations even if the radiometer is slightly
unbalanced.

In LFI baseline we foresee a software implementation of the gain
modulation factor $r$, so that the sky and reference load signals
will be detected, converted to digital and then subtracted after
multiplication of the reference load temperature by $r$.

Although $r$ will not suffer any fluctuations, it will be in
general different, at any time, from the {\sl ideal} value $r^*$.
This has an impact on the radiometer $1/f$ noise at two different
levels: (i) $f_k^0(G)$ will be different from zero (because the
effect of gain oscillations will not be cancelled completely) and
(ii) $f_k^0(T_n)$ will decrease or increase depending on the sign
of $\epsilon_r$ (because $r$ will be closer or farther from 1).

In order to evaluate the effect of a slight deviation of $r$ from
$r^*$ let us consider a reference value of $r$ that nulls or
makes very close to zero the output at a certain instant $t_0$
(i.e. $r(t_0)=r^*(t_0)$); therefore at a generic time $t$ the
parameter $r(t)$ can be written as:
$r(t)=r^*(t_0)(1+\epsilon_r(t))$, where $\epsilon_r(t)\ll 1$.

Considering that for $r\simeq r^*$ we have $f_k^0(G)\simeq
\frac{1}{4}\,C^2
\beta\frac{\left[T_x+T_n-r(T_y+T_n)\right]^2}{(T_x+T_n)^2} $ (see
equation (\ref{eq:kneefreq_G})) we obtain the
following expressions for $f_k(T_n)$ and $f_k(G)$:

\begin{eqnarray}
&& f_k^0(T_n)=\beta \,
   {\left[ \frac{A\,{T_n}}
       {2\,\left( {T_n} + {T_x} \right) } \right] }^2\,
   {\left[ 1 - \frac{\left( 1 + {{\epsilon }_r} \right) \,
          \left( {T_n} + {T_x} \right) }{{T_n} + {T_y}}
       \right] }^2\nonumber\\
&&\mbox{}\\
&& f_k^0(G)=\beta \,{N_s}\,
   {\left\{ \frac{A\,\left[ {T_n} + {T_x} -
           \left( 1 + {{\epsilon }_r} \right) \,
            \left( {T_n} + {T_x} \right)  \right] }{{T_
           n} + {T_x}} \right\} }^2\nonumber
\label{eq:fk_r!=r*}
\end{eqnarray}

If we solve for $\epsilon_r$ the equation $f_k^0(T_n)=f_k^0(G)$
we find two solutions, $\epsilon_{r_1}$ and $\epsilon_{r_2}$
(with $\epsilon_{r_1}\,<\,\epsilon_{r_2}$) given by:
\begin{eqnarray}
\label{eq:epsr1_epsr2}
&& \epsilon_{r_1}=-\frac{{T_n}\,
     \left( {T_y} - {T_x} \right) }{\left( {T_n} +
       {T_x} \right) }\,\left[ \left(
          C/A-1 \right) \,{T_n} +
       (C/A)\,{T_y} \right]^{-1} \nonumber\\
&&\mbox{}\\
&& \epsilon_{r_2}=\frac{{T_n}\,
       \left({T_y} -{T_x} \right) }{{T_n} + {T_x}}
     \left[\left( C/A + 1 \right) \,{T_n} +
     (C/A)\,{T_y}\right]^{-1}\nonumber
\end{eqnarray}

\noindent which define an interval
$\left[\epsilon_{r_1},\epsilon_{r_2}\right]$ such that $f_k(G)
\ll f_k(T_n)$ for $\epsilon_{r_1}\ll\epsilon_r\ll\epsilon_{r_2}$
and $f_k(G) \gg f_k(T_n)$ otherwise. From equation
(\ref{eq:epsr1_epsr2}) it is apparent that the width of the
interval $\left[\epsilon_{r_1},\epsilon_{r_2}\right]$ is smaller
for the high frequency channels, characterised by higher values of
the noise temperature; therefore the requirement on the gain
modulation factor accuracy is determined by the 100~GHz channel.

\begin{figure}[here]
\begin{center}
\resizebox{9 cm}{!}{\includegraphics{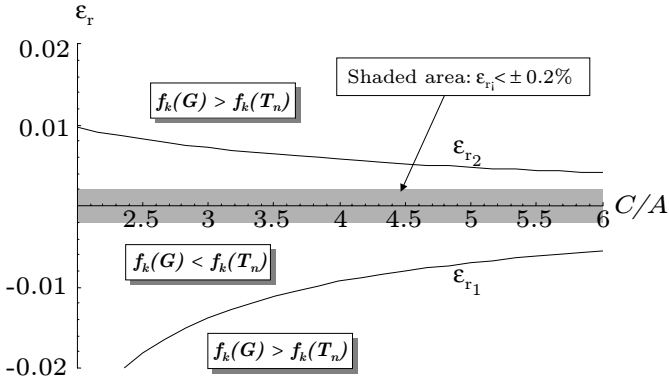}}
\end{center}
\caption{Behaviour of $\epsilon_{r_1}$ and $\epsilon_{r_2}$
(calculated from equation (\ref{eq:epsr1_epsr2})) as a function of
$C/A$ for the 100~GHz channel. From the figure if follows that
$f_k(G)< f_k(T_n)$ (i.e. noise temperature fluctuations dominate)
if the gain modulation factor is stable at the level of
$\leq\pm\, 0.2\,\%$.\label{fig:eps_r}}
\end{figure}

Figure \ref{fig:eps_r} shows the behaviour of $\epsilon_{r_1}$ and
$\epsilon_{r_2}$ versus the ratio between gain and noise
temperature fluctuations (i.e. $C/A=2\,\sqrt{N_s}$) for the 100
GHz channel. The region between the two curves corresponds to
values of $\epsilon_r$ for which the noise temperature $1/f$
fluctuations dominate over gain instabilities. From the figure it
is apparent that $1/f$ noise arising from gain fluctuations do
not dominate if the gain modulation factor is accurate at the
level of $\leq\pm\,0.2\,\%$.

Let us estimate the uncertainty in $r$ introduced by the largest
expected fluctuation in the signal, i.e. the CMB dipole. If the
fluctuation induced by the Dipole were larger than the
requirement, then it would be necessary to calculate $r$ almost in
real-time, in sub-minute time scales. From a simple calculation
of $\delta r/r = \delta T_{\rm dipole}/ \left(T_x+T_n\right)$ it
follows that $\delta r / r \sim 10^{-4}$, which indicates that it
will be possible to update $r$ on longer time-scales (of the order
of few days) to account for variations caused by slow
instrumental drifts.

Let us now analyse the effect of a small leakage across the first
hybrid coupler, in the balanced radiometer approximation; if
$\epsilon$ is the fraction of a signal applied to one port that
shows up in the isolated port, then we can write the averaged
power output, $\overline p$, as:

\begin{equation}
\overline{p} = {a k \beta G}\left\{ { T_x + T_n { (1+
\epsilon)}^2 - r \left[{T_y + T_n { (1- \epsilon)}^2}\right]
}\right\}
\end{equation}

The change in output caused by a small change in gain with the
presence of such non-ideal isolation is:

\begin{equation}
\frac{\partial \overline{p} }{\partial G} = {a k \beta }\left\{ {
T_x + T_n { (1+ \epsilon)}^2 - r \left[{T_y + T_n { (1-
\epsilon)}^2}\right] }\right\}
\end{equation}

Now, with the correct choice of $r$, we can make the sensitivity
to gain fluctuations vanish and null the average output of the
radiometer simultaneously:

\begin{equation}
r = \frac{T_x + T_n { (1+ \epsilon)}^2 } {T_y + T_n { (1-
\epsilon)}^2}
\end{equation}

Note that this freedom of choice for $r$ is not available to
correlation radiometer designs without gain modulation, and
therefore they are subject to $1/f$ fluctuations on the
non-isolated fraction of the signal.

\subsection{Sensitivity to back-end amplifier gain fluctuations\label{sec:sensitivity_BE_amps}}

In this section we analyse the radiometer sensitivity to gain
fluctuations of the back end amplifiers. Considering that the
back-end noise is suppressed by a factor $\sim 10^3$ by the
front-end amplification we can neglect the contribution of the
noise temperature fluctuations.

Let us consider the case of a perfectly balanced front-end and
suppose that the gain value in the two states of the phase switch
is different by $\Delta G_B$.

The power output from a single
radiometer leg will be then given by:

\begin{eqnarray}
&& p = p_x - r\, p_y\nonumber\\
&&\mbox{}\nonumber\\
&& p_{x}=a\, k\, \beta \, G_F \,
G_B\left(T_x+T_{n_F}+T_{n_B}/G_F\right)\\
&&\mbox{}\nonumber\\
&& p_{y}=a\, k\, \beta \, G_F\,( G_B+\Delta
G_B)\left(T_x+T_{n_F}+T_{n_B}/G_F\right)\nonumber
\end{eqnarray}

A straightforward analysis shows that the fluctuation $\Delta
G_B$ will cause an equivalent spurious signal

\begin{equation}
\Delta T_{\rm eq}= -r\,\frac{\Delta
G_B}{G_B}\left(T_y+T_{n_F}+T_{n_B}/G_F\right)
\end{equation}

Substituting $\Delta G_B/G_B = C/\sqrt{f}$, $r=r^*_0$ and setting
$\Delta T_{\rm equiv}$ equal to the single radiometer leg
spectral density (equal to $2\,
\beta^{-1/2}\left(T_x+T_{n_{tot}}\right)$, see equation
(\ref{eq:basic_case_approx})) we can calculate the post-detection
knee-frequency as:

\begin{equation}
f_k^0(G_B)=\beta \,N_s\,A^2 \label{eq:backend_fk}
\end{equation}

Using typical values for $\beta$, $N_s$ and $A$ in equation
(\ref{eq:backend_fk}) it follows that the post-detection knee
frequency arising from back-end gain fluctuations can be much
greater than 1~Hz. Therefore to eliminate the contribution of the
back-end $1/f$ noise we need a phase switch frequency such that
$f_{\rm sw}\gg f_k^0(G_B)$, so that the gain will be constant
during the post-detection integration time.

For the LFI radiometers we have chosen a phase switch frequency of
$\sim$ 4~KHz which leads to a small back-end $1/f$ contribution
to the overall radiometer knee frequency for expected values of
back-end gain fluctuations.

\subsection{Sensitivity to reference load fluctuations\label{sec:sensitivity_reference_load}}

Proceeding as in the previous section we have that the signal
mimicked by reference load fluctuations, $\Delta T_y$, is given by 
$\Delta T_{\rm eq} = - r \Delta T_y$.

If we consider random fluctuations in the reference load signal with 
a spectral density given by $\Delta T_y(f)$ then we have that these fluctuations
equal the white noise for:

\begin{equation}
\Delta T_y(f)  =\Delta T_y^0(f)\left[1+\frac{\alpha_3 \,
\epsilon_{T_n}}{4}
-\left(r-\frac{1}{r}\right)\frac{\epsilon_G^2}{16}\right]
\label{eq:tyeqn}
\end{equation}

\noindent where $\Delta T_y^0(f)= \sqrt{2/\beta}\,(T_y+T_n)$ and
$\alpha_3$ is defined in equation (\ref{eq:alpha_1_to_4}).

This provides an upper limit to allowed random fluctuations in $T_y$ 
in order not to dominate the noise of
the radiometer. For typical LFI parameters, this upper limit 
is of the order of $\Delta T_y\sim 1.8\times 10^{-4} {\rm
K}/\sqrt{\rm Hz}$ at 30~GHz, slightly increasing for the higher
frequency channels ($3.2\times 10^{-4} {\rm K}/\sqrt{\rm Hz}$ at
100~GHz). If we now consider the reference load in a global systematic
error budget for LFI (the details of which are outside the scope of our 
paper) this implies requirements that are about one order of magnitude
smaller than the above values.

If we consider non random fluctuations in the reference signal then 
the error induced by these variations must be much less than the 
sensitivity per pixel in the final maps. This error, that depends on 
the spectral behaviour of the spurious signal, on the satellite scanning 
strategy and on the data analysis procedures to build the sky maps, can 
be estimated using the approach described in Mennella et al. 
(\cite{mennella2002}). If we follow this approach considering that 
the parameter $r$ is of order unity, 
we have that spin synchronous variations
of the reference signal will be transferred directly to the final maps,
while slower fluctuations will be damped by a factor of the order of
$10^{-2}$~-~$10^{-3}$ by the measurement strategy and by data analysis.

\section{Conclusions\label{sec:conclusions}}

In this paper we have discussed the pseudo-correlation
architecture adopted for the radiometers of the {\sc Planck}-LFI
instrument and we have studied the sensitivity of the measured
signal to various systematic effects. In our treatment we have
considered both the ideal case of a perfectly balanced radiometer
and the effect of small mismatches in the various radiometer
parameters.

The first result is that the radiometer sensitivity does not
depend on the level of the reference load temperature; even in
the case of a slight imbalance in the radiometer parameters the
dependence on $T_y$ is at the level of $\partial \Delta T
/\partial T_y\sim 10^{-5}$ which is negligible. The only mismatch
which has a first-order impact on $\Delta T$ is the noise
temperature mismatch of the front-end amplifiers; our analysis
shows that it is possible to maintain the sensitivity degradation
below 1\% with a noise temperature match better than 5\%.

With proper gain modulation ($r=r^*_0$) the $1/f$ noise in the
radiometer output is determined mainly by noise temperature
fluctuations in the front-end amplifiers, with a knee frequency of
few mHz, provided that the front-end amplifier amplitude match is
better that $\sim\pm 0.5$~dB. Such a high level of $1/f$ noise
suppression depends on the gain modulation factor, which must be
determined with an accuracy better than $\pm 0.2\%$.  If the
accuracy on $r$ is less than $\pm 1\%$ then gain fluctuations become
the major source of $1/f$ noise; if $\epsilon_r$ is kept in the
range $\pm 1\%$ we expect values of the knee frequency of the
order of 50~mHz which can be easily handled by destriping
algorithms. The presence of a small amount of leakage in the
first hybrid does not significantly modify this conclusion; with
the correct choice of $r$, one can make the sensitivity to gain
fluctuations vanish and null the average output of the radiometer
simultaneously.

The effect of gain fluctuations in the back-end amplifiers can be
made negligible by the fast front-end switching between sky and
reference load signals. The LFI baseline of 4096~Hz for the phase
switch frequency has been chosen to guarantee a high suppression
level of the $1/f$ noise from back-end amplifiers.

In general our analysis demonstrates the effectiveness of the
gain modulation concept applied to this form of radiometer. The
estimate of the knee frequency given by equation
(\ref{eq:kneefreq}) is quite low and relatively immune to small
imperfections in radiometer balance. The modified correlation
radiometer scheme reduces the knee frequency by more than two
orders of magnitude, compared to a total power radiometer of
similar bandwidth and intrinsic transistor fluctuations. For such
small residual knee frequency (of $\sim 0.1$~Hz) it will be
possible to remove efficiently the effects during data analysis.

We have also studied the sensitivity of the radimeter to changes in the
reference load temperature, $T_y$. Our analysis provides a framework in
which thermal stability requirements on the LFI reference loads can be
evaluated.

A refinement of the present analysis for the determination of
$f_k$ will be pursued in the future by software simulations of
the radiometer functions to accurately study the combined effect
of all components. Finally, laboratory measurements of a
prototype radiometer working under conditions close to those of
Planck mission constitute the most important checks for the
ultimately understanding of the behaviour of {\sc Planck}~LFI
radiometers regarding the $1/f$ type noise and possible further
effects. Results of preliminary laboratory measurements performed
of {\sc Planck}-LFI prototype radiometers will be presented in
forthcoming publications.

\begin{acknowledgements}
It is a pleasure to thank S. Weinreb, T. Gaier, D. Scott, G.
Smoot, C. Lawrence, S. Levin, M. Janssen, and J. Delabrouille for
useful discussions. \end{acknowledgements}

\appendix

\onecolumn
\section{\label{app:power_output}Power output}

The radiometer power output at each of the radiometer output legs
is defined by:

\begin{eqnarray}
 p_1(t)= && \frac{a}{4}\,g_{B_1}^2\,\left\{\left[
 \left(2\, n_{B_1} \cos{(\phi_{B_1})}+g_{F_1}
 (\sqrt{2}\,n_{F_1}+x+y)\cos{(\phi_{F_1}+\phi_{B_1})}+
 \right.\right.\right.\nonumber\\
&&\,\,\,\,\,\,\,\,\,\,\,\,\,\,\,\,\,\,\,\,\,\,\,\,
\left. +\sqrt{A_1}\,g_{F_2}(\sqrt{2}\,n_{F_2}+x-y)
\cos{(\theta_1+\phi_{F_2}+\phi_{B_1})}\right)^2+\nonumber\\
 &&+\left(2\, n_{B_1} \sin{(\phi_{B_1})}+g_{F_1}
 (\sqrt{2}\,n_{F_1}+x+y)\sin{(\phi_{F_1}+\phi_{B_1})}+\right.\nonumber\\
&&\,\,\,\,\,\,\,\,\,\,\,\,\,\,\,\,\,\,\,\,\,\,\,\,
\left.\left.+\sqrt{A_1}\,g_{F_2}(\sqrt{2}\,n_{F_2}+x-y)
 \sin{(\theta_1+\phi_{F_2}+\phi_{B_1})}\right)^2\right]+\nonumber\\
 &&-r\,\left[\left(2\, n_{B_1} \cos{(\phi_{B_1})}+
 g_{F_1}(\sqrt{2}\,n_{F_1}+x+y)\cos{(\phi_{F_1}+\phi_{B_1})}+
 \right.\right.\nonumber\\
&&\,\,\,\,\,\,\,\,\,\,\,\,\,\,\,\,\,\,\,\,\,\,\,\,
\left.+\sqrt{A_2}\,g_{F_2}(\sqrt{2}\,n_{F_2}+x-y)
\cos{(\theta_2+\phi_{F_2}+\phi_{B_1})}\right)^2+\nonumber\\
 &&+\left(2\,n_{B_1} \sin{(\phi_{B_1})}+g_{F_1}
 (\sqrt{2}\,n_{F_1}+x+y)\sin{(\phi_{F_1}+\phi_{B_1})}+\right.\nonumber\\
&&\,\,\,\,\,\,\,\,\,\,\,\,\,\,\,\,\,\,\,\,\,\,\,\,
 \left.\left.\left.+\sqrt{A_2}\,g_{F_2}(\sqrt{2}\,n_{F_2}+x-y)\sin{(\theta_2+\phi_{F_2}+\phi_{B_1})}\right)^2\right]\right\}
    \label{eq:p(t)}
\end{eqnarray}

\noindent where $a$ is the constant of proportionality of the
square-law detectors. A similar relationship holds for $p_2(t)$.

\section{\label{app:sensitivity}Sensitivity}

\subsection{Outline for sensitivity calculation}

In this appendix we outline the procedure to calculate the LFI
radiometer sensitivity (see equation (\ref{eq:sensitivity})). We
start from the expression of the power output $p(t)$, which is
given by equation (\ref{eq:p(t)}) and then we calculate the
autocorrelation of $p(t)$, denoted as $\psi_{p}({\tau}) \equiv
\overline{p(t)p(t+\tau)}$. Note that we assume that $x$, $y$,
$n_1$, and $n_2$ are all uncorrelated gaussian variables so that
$\overline{xy} = \overline{x} \, \overline{y}$ and that they all
have zero mean. The next step is the calculation of the Fourier
transform of $\psi_p(\tau)$ (indicated by $Q_p(f)$, which can be
split in a constant part, $Q_p(0)$, and in a fluctuating part
$Q_p'(f)$. Then we calculate the rms voltage $w_{\rm
rms}={\overline{w(t)^2}}^{1/2}$ where
$\overline{w(t)^2}=\int_{-\infty}^{+\infty}df H(f)Q_p'(f)$ and
$H(f)$ is a rectangular lowpass filter of height $H_0$ and width
$b$. The last step is to calculate the change in power output
$\Delta p$ determined by a change in the sky signal equal to
$\Delta T$; the sensitivity $\Delta T$ is then calculated by
solving the equation $w_{\rm rms}=H_0^{1/2}\Delta p$.

\subsection{Sensitivity analytical form}

The radiometer sensitivity $\Delta T$ (calculated according to
the procedure outlined above) has the following form:

\begin{equation}
\Delta  T = \frac{1}{2}\left[ \Delta  {T_1^2} + \Delta
{T_2^2}\right]^{\frac{1}{2}} \label{eq:sensitivity}
\end{equation}

\noindent where $\Delta T_1$ and $\Delta T_2$ (which represent
the minimum detectable rms signal on each single radiometer leg)
are defined by the following equations:

\begin{eqnarray}
\Delta  {T_1} =
   {\sqrt{\frac{1}{2 \beta  \tau }}}
    \frac{{\left( A - B\, r + C\, r^2  \right) }^{\frac{1}{2}}}
     {\hat G - r \hat I}\nonumber\\
\Delta  {T_2} =
   {\sqrt{\frac{1}{2 \beta  \tau }}}
    \frac{{\left( D - E\, r + F\, r^2  \right) }^{\frac{1}{2}}}
     {\tilde G - r \tilde I}
\label{eq:DT3_DT4}
\end{eqnarray}

The coefficients $A$ through $F$ have the following form:

\begin{equation}
A = a_1 T_x^2 + a_2 T_y^2 + a_3 T_x +
    a_4 T_y + a_5 T_x T_y + a_6
\end{equation}

\noindent with similar relationships for $B$, $C$, $D$, $E$ and
$F$. Coefficients $a_j$, $b_j$, etc. are defined as follows:

\subsubsection{Definitions of $a_j$
coefficients\label{app:aj_coefficients}}

\begin{eqnarray}
a_1 = && 2 \,\hat G^2\nonumber\\
a_2 = && 2 \,\tilde I^2\nonumber\\
a_3 = && a_{31}\,T_{n_{F_1}}+a_{32}\,T_{n_{F_2}}+a_{33}\,T_{n_{B_1}}\nonumber\\
&& {a_{31}} =
  \frac{1}{2}\,{G_{F_1}}\,
   {\left[ {\sqrt{{G_{F_1}}}} + {\sqrt{{A_1}\,{G_{F_2}}}}\,
        \cos ({{\theta }_1} - {{\phi }_{F_1}} + {{\phi }_{F_2}}) \right] }^2 \nonumber\\
&&{a_{32}} =
   \frac{1}{2}\,{A_1}\,{G_{F_2}}\,
   {\left[ {\sqrt{{A_1}\,{G_{F_2}}}} + {\sqrt{{G_{F_1}}}}\,
        \cos ({{\theta }_1} - {{\phi }_{F_1}} + {{\phi }_{F_2}}) \right] }^2\nonumber\\
&& {a_{33}} =
   {\left[ {\sqrt{{G_{F_1}}}}\,\cos ({{\phi }_{F_1}}) +
      {\sqrt{{A_1}\,{G_{F_2}}}}\,\cos ({{\theta }_1} + {{\phi }_{F_2}}) \right] }^2 \nonumber\\
a_4 = && a_{41}\,{T_{{n_{F_1}}}}+{a_{42}}\,{T_{{n_2}}}+
{a_{43}}\,{T_{{n_{B_1}}}}\nonumber\\
&& {a_{41}} =
   \frac{1}{2}\,{G_{F_1}}\,
   {\left[ {\sqrt{{G_{F_1}}}} - {\sqrt{{A_1}\,{G_{F_2}}}}\,
        \cos ({{\theta }_1} - {{\phi }_{F_1}} + {{\phi }_{F_2}}) \right] }^2\nonumber\\
&& {a_{42}} =
   \frac{1}{2}\,{A_1}\,{G_{F_2}}\,
   {\left[ {\sqrt{{A_1}\,{G_{F_2}}}} - {\sqrt{{G_{F_1}}}}\,
        \cos ({{\theta }_1} - {{\phi }_{F_1}} + {{\phi }_{F_2}}) \right] }^2\nonumber\\
&&{a_{43}} =
  {\left[ {\sqrt{{G_{F_1}}}}\,\cos ({{\phi }_{F_1}}) -
      {\sqrt{{A_1}\,{G_{F_2}}}}\,\cos ({{\theta }_1} + {{\phi }_{F_2}}) \right] }^2\nonumber\\
{a_5} = &&
   \frac{1}{4}\,{\left( {G_{F_1}} - {A_1}\,{G_{F_2}} \right)
   }^2\nonumber\\
{a_6} = &&
   \frac{1}{2}\,{G_{F_1}}\,{T_{{n_{F_1}}}}\,
     \left[ {G_{F_1}}\,{T_{{n_{F_1}}}} +
       4\,{\cos ({{\phi }_{F_1}})}^2\,{T_{{n_{B_1}}}} \right]
       +\nonumber\\
&&+ \frac{1}{2}\,{A_1}\,{G_{F_2}}\,{T_{{n_{F_2}}}}\,
     \left[ 2\,{\cos ({{\theta }_1} - {{\phi }_{F_1}} +
            {{\phi }_{F_2}})}^2\,{G_{F_1}}\,{T_{{n_{F_1}}}} +
       {A_1}\,{G_{F_2}}\,{T_{{n_2}}} +
       4\,{\cos ({{\theta }_1} + {{\phi }_{F_2}})}^2\,{T_{{n_{B_1}}}}
       \right]  + 2\,{{T_{{n_{B_1}}}^2}}\nonumber
\end{eqnarray}

\subsubsection{Definitions of $b_j$
coefficients\label{app:bj_coefficients}}

\begin{eqnarray}
b_1 = && 4 \,\hat I \, \hat G\nonumber\\
b_2 = && 4 \,\tilde I \, \tilde G\nonumber\\
b_3 = && b_{31}\,T_{n_{F_1}}+b_{32}\,T_{n_{F_2}}+b_{33}\,T_{n_{B_1}}\nonumber\\
&& {b_{31}} =
   {G_{F_1}}\,\left[ {\sqrt{{G_{F_1}}}} +
     {\sqrt{{A_1}\,{G_{F_2}}}}\,\cos ({{\theta }_1} - {{\phi }_{F_1}} + {{\phi }_{F_2}}) \right] \,
   \left[ {\sqrt{{G_{F_1}}}} + {\sqrt{{A_2}\,{G_{F_2}}}}\,
      \cos ({{\theta }_2} - {{\phi }_{F_1}} + {{\phi }_{F_2}}) \right] \nonumber\\
&&{b_{32}} =
 {\sqrt{{A_1}\,{A_2}}}\,{G_{F_2}}\,
   \left[ {\sqrt{{A_1}\,{G_{F_2}}}} + {\sqrt{{G_{F_1}}}}\,
      \cos ({{\theta }_1} - {{\phi }_{F_1}} + {{\phi }_{F_2}}) \right] \,
   \left[ {\sqrt{{A_2}\,{G_{F_2}}}} + {\sqrt{{G_{F_1}}}}\,
      \cos ({{\theta }_2} - {{\phi }_{F_1}} + {{\phi }_{F_2}}) \right] \nonumber\\
&&{b_{33}} =
  2\,\left[ {\sqrt{{G_{F_1}}}}\,\cos ({{\phi }_{F_1}}) +
     {\sqrt{{A_1}\,{G_{F_2}}}}\,\cos ({{\theta }_1} + {{\phi }_{F_2}}) \right] \,
   \left[ {\sqrt{{G_{F_1}}}}\,\cos ({{\phi }_{F_1}}) +
     {\sqrt{{A_2}\,{G_{F_2}}}}\,\cos ({{\theta }_2} + {{\phi }_{F_2}}) \right] \nonumber\\
b_4 = && b_{41}\,{T_{{n_{F_1}}}}+{b_{42}}\,{T_{{n_{F_2}}}}+
{b_{43}}\,{T_{{n_{B_1}}}}\nonumber\\
&& {b_{41}} =
   {G_{F_1}}\,\left[ {\sqrt{{G_{F_1}}}} -
     {\sqrt{{A_1}\,{G_{F_2}}}}\,\cos ({{\theta }_1} - {{\phi }_{F_1}} + {{\phi }_{F_2}}) \right] \,
   \left[ {\sqrt{{G_{F_1}}}} - {\sqrt{{A_2}\,{G_{F_2}}}}\,
      \cos ({{\theta }_2} - {{\phi }_{F_1}} + {{\phi }_{F_2}}) \right] \nonumber\\
&&{b_{42}} =
 {\sqrt{{A_1}\,{A_2}}}\,{G_{F_2}}\,
   \left[ {\sqrt{{A_1}\,{G_{F_2}}}} - {\sqrt{{G_{F_1}}}}\,
      \cos ({{\theta }_1} - {{\phi }_{F_1}} + {{\phi }_{F_2}}) \right] \,
   \left[ {\sqrt{{A_2}\,{G_{F_2}}}} - {\sqrt{{G_{F_1}}}}\,
      \cos ({{\theta }_2} - {{\phi }_{F_1}} + {{\phi }_{F_2}}) \right] \nonumber\\
&&{b_{43}} =
  2\,\left[ {\sqrt{{G_{F_1}}}}\,\cos ({{\phi }_{F_1}}) -
     {\sqrt{{A_1}\,{G_{F_2}}}}\,\cos ({{\theta }_1} + {{\phi }_{F_2}}) \right] \,
   \left[ {\sqrt{{G_{F_1}}}}\,\cos ({{\phi }_{F_1}}) -
     {\sqrt{{A_2}\,{G_{F_2}}}}\,\cos ({{\theta }_2} + {{\phi }_{F_2}}) \right] \nonumber\\
{b_5} =&&
   \frac{1}{2}\,\left( {G_{F_1}} - {A_1}\,{G_{F_2}} \right) \,
    \left( {G_{F_1}} - {A_2}\,{G_{F_2}} \right)\nonumber\\
{b_6} =&&
   {G_{F_1}}\,{T_{{n_{F_1}}}}\,\left[ {G_{F_1}}\,{T_{{n_{F_1}}}} +
4\,{\cos ({{\phi }_{F_1}})}^2\,{T_{{n_{B_1}}}} \right]+  \nonumber\\
+&&{G_{F_2}}\,\sqrt{{A_1}\,{A_2}}{T_{{n_{F_2}}}}\,\left[
\sqrt{{A_1}\,{A_2}}\,{G_{F_2}}\,
        {T_{{n_{F_2}}}} + 4\,{T_{{n_{B_1}}}}\,
        \cos ({{\theta }_1} + {{\phi }_{F_2}})\,
        \cos ({{\theta }_2} + {{\phi }_{F_2}}) +\right.\nonumber\\
+&&\left.2\,{G_{F_1}}\,{T_{{n_{F_1}}}}\,
        \cos ({{\theta }_1} - {{\phi }_{F_1}} + {{\phi }_{F_2}})\,
        \cos ({{\theta }_2} - {{\phi }_{F_1}} + {{\phi }_{F_2}}) \right]
     + 4\,{{T_{{n_{B_1}}}^2}}\nonumber
\end{eqnarray}

\subsubsection{Definitions of $c_j$
coefficients\label{app:cj_coefficients}}

\begin{eqnarray}
c_1 = && 2 \,\hat I^2\nonumber\\
c_2 = && 2 \,\tilde G^2\nonumber\\
c_3 = && c_{31}\,T_{n_{F_1}}+c_{32}\,T_{n_{F_2}}+c_{33}\,T_{n_{B_1}}\nonumber\\
&& {c_{31}} =
  \frac{1}{2}\,{G_{F_1}}\,
   {\left[ {\sqrt{{G_{F_1}}}} + {\sqrt{{A_2}\,{G_{F_2}}}}\,
        \cos ({{\theta }_2} - {{\phi}_{F_1}} + {{\phi}_{F_2}}) \right] }^2 \nonumber\\
&&{c_{32}} =
   \frac{1}{2}\,{A_2}\,{G_{F_2}}\,
   {\left[ {\sqrt{{A_2}\,{G_{F_2}}}} + {\sqrt{{G_{F_1}}}}\,
        \cos ({{\theta }_2} - {{\phi}_{F_1}} + {{\phi}_{F_2}}) \right] }^2\nonumber\\
&& {c_{33}} =
   {\left[ {\sqrt{{G_{F_1}}}}\,\cos ({{\phi}_{F_1}}) +
      {\sqrt{{A_2}\,{G_{F_2}}}}\,\cos ({{\theta }_2} + {{\phi}_{F_2}}) \right] }^2 \nonumber\\
c_4 = && c_{41}\,{T_{{n_{F_1}}}}+{c_{42}}\,{T_{{n_{F_2}}}}+
{c_{43}}\,{T_{{n_{B_1}}}}\nonumber\\
&& {c_{41}} =
   \frac{1}{2}\,{G_{F_1}}\,
   {\left[ {\sqrt{{G_{F_1}}}} - {\sqrt{{A_2}\,{G_{F_2}}}}\,
        \cos ({{\theta }_2} - {{\phi}_{F_1}} + {{\phi}_{F_2}}) \right] }^2\nonumber\\
&& {c_{42}} =
   \frac{1}{2}\,{A_2}\,{G_{F_2}}\,
   {\left[ {\sqrt{{A_2}\,{G_{F_2}}}} - {\sqrt{{G_{F_1}}}}\,
        \cos ({{\theta }_2} - {{\phi}_{F_1}} + {{\phi}_{F_2}}) \right] }^2\nonumber\\
&&{c_{43}} =
  {\left[ {\sqrt{{G_{F_1}}}}\,\cos ({{\phi}_{F_1}}) -
      {\sqrt{{A_2}\,{G_{F_2}}}}\,\cos ({{\theta }_2} + {{\phi}_{F_2}}) \right] }^2\nonumber\\
{c_5} = &&
   \frac{1}{4}\,{\left( {G_{F_1}} - {A_2}\,{G_{F_2}} \right)
   }^2\nonumber\\
{c_6} = &&
   \frac{1}{2}\,{G_{F_1}}\,{T_{{n_{F_1}}}}\,
     \left[ {G_{F_1}}\,{T_{{n_{F_1}}}} +
       4\,{\cos ({{\phi}_{F_1}})}^2\,{T_{{n_{B_1}}}} \right]
       +\nonumber\\
&&+ \frac{1}{2}\,{A_2}\,{G_{F_2}}\,{T_{{n_{F_2}}}}\,
     \left[ 2\,{\cos ({{\theta }_2} - {{\phi}_{F_1}} +
            {{\phi}_{F_2}})}^2\,{G_{F_1}}\,{T_{{n_{F_1}}}} +
       {A_2}\,{G_{F_2}}\,{T_{{n_{F_2}}}} +
       4\,{\cos ({{\theta }_2} + {{\phi}_{F_2}})}^2\,{T_{{n_{B_1}}}}
       \right]  + 2\,{{T_{{n_{B_1}}}^2}}\nonumber
\end{eqnarray}

\subsubsection{Definitions of $d_j$
coefficients\label{app:dj_coefficients}}

\begin{eqnarray}
d_1 = && 2 \,\tilde G^2\nonumber\\
d_2 = && 2 \,\hat I^2\nonumber\\
d_3 = && d_{3,1}\,T_{n_{F_1}}+d_{3,2}\,T_{n_{F_2}}+d_{3,3}\,T_{n_{B_2}}\nonumber\\
&& {d_{3,1}} =
  \frac{1}{2}\,{G_{F_1}}\,
   {\left[ {\sqrt{{G_{F_1}}}} - {\sqrt{{A_2}\,{G_{F_2}}}}\,
        \cos ({{\theta }_2} - {{\phi}_{F_1}} + {{\phi}_{F_2}}) \right] }^2 \nonumber\\
&&{d_{3,2}} =
   \frac{1}{2}\,{A_2}\,{G_{F_2}}\,
   {\left[ {\sqrt{{A_2}\,{G_{F_2}}}} - {\sqrt{{G_{F_1}}}}\,
        \cos ({{\theta }_2} - {{\phi}_{F_1}} + {{\phi}_{F_2}}) \right] }^2\nonumber\\
&& {d_{3,3}} =
   {\left[ {\sqrt{{G_{F_1}}}}\,\cos ({{\phi}_{F_1}}) -
      {\sqrt{{A_2}\,{G_{F_2}}}}\,\cos ({{\theta }_2} + {{\phi}_{F_2}}) \right] }^2 \nonumber\\
d_4 = && d_{4,1}\,{T_{{n_{F_1}}}}+{d_{4,2}}\,{T_{{n_{F_2}}}}+
{d_{4,3}}\,{T_{{n_{B_2}}}}\nonumber\\
&& {d_{4,1}} =
   \frac{1}{2}\,{G_{F_1}}\,
   {\left[ {\sqrt{{G_{F_1}}}} + {\sqrt{{A_2}\,{G_{F_2}}}}\,
        \cos ({{\theta }_2} - {{\phi}_{F_1}} + {{\phi}_{F_2}}) \right] }^2\nonumber\\
&& {d_{4,2}} =
   \frac{1}{2}\,{A_2}\,{G_{F_2}}\,
   {\left[ {\sqrt{{A_2}\,{G_{F_2}}}} + {\sqrt{{G_{F_1}}}}\,
        \cos ({{\theta }_2} - {{\phi}_{F_1}} + {{\phi}_{F_2}}) \right] }^2\nonumber\\
&&{d_{4,3}} =
  {\left[ {\sqrt{{G_{F_1}}}}\,\cos ({{\phi}_{F_1}}) +
      {\sqrt{{A_2}\,{G_{F_2}}}}\,\cos ({{\theta }_2} + {{\phi}_{F_2}}) \right] }^2\nonumber\\
{d_5} = &&
   \frac{1}{4}\,{\left( {G_{F_1}} - {A_2}\,{G_{F_2}} \right)
   }^2\nonumber\\
{d_6} = &&
   \frac{1}{2}\,{G_{F_1}}\,{T_{{n_{F_1}}}}\,
     \left[ {G_{F_1}}\,{T_{{n_{F_1}}}} +
       4\,{\cos ({{\phi}_{F_1}})}^2\,{T_{{n_{B_2}}}} \right]
       +\nonumber\\
&&+ \frac{1}{2}\,{A_2}\,{G_{F_2}}\,{T_{{n_{F_2}}}}\,
     \left[ 2\,{\cos ({{\theta }_2} - {{\phi}_{F_1}} +
            {{\phi}_{F_2}})}^2\,{G_{F_1}}\,{T_{{n_{F_1}}}} +
       {A_2}\,{G_{F_2}}\,{T_{{n_{F_2}}}} +
       4\,{\cos ({{\theta }_2} + {{\phi}_{F_2}})}^2\,{T_{{n_{B_2}}}}
       \right]  + 2\,{{T_{{n_{B_2}}}^2}}\nonumber
\end{eqnarray}

\subsubsection{Definitions of $e_j$
coefficients\label{app:ej_coefficients}}

\begin{eqnarray}
e_1 = && 4 \,\tilde I \, \tilde G\nonumber\\
e_2 = && 4 \,\hat I \, \hat G\nonumber\\
e_3 = && d_{3,1}\,T_{n_{F_1}}+d_{3,2}\,T_{n_{F_2}}+d_{3,3}\,T_{n_{B_2}}\nonumber\\
&& {e_{3,1}} =
   {G_{F_1}}\,\left[ {\sqrt{{G_{F_1}}}} -
     {\sqrt{{A_1}\,{G_{F_2}}}}\,\cos ({{\theta }_1} - {{\phi}_{F_1}} + {{\phi}_{F_2}}) \right] \,
   \left[ {\sqrt{{G_{F_1}}}} - {\sqrt{{A_2}\,{G_{F_2}}}}\,
      \cos ({{\theta }_2} - {{\phi}_{F_1}} + {{\phi}_{F_2}}) \right] \nonumber\\
&&{e_{3,2}} =
 {\sqrt{{A_1}\,{A_2}}}\,{G_{F_2}}\,
   \left[ {\sqrt{{A_1}\,{G_{F_2}}}} - {\sqrt{{G_{F_1}}}}\,
      \cos ({{\theta }_1} - {{\phi}_{F_1}} + {{\phi}_{F_2}}) \right] \,
   \left[ {\sqrt{{A_2}\,{G_{F_2}}}} - {\sqrt{{G_{F_1}}}}\,
      \cos ({{\theta }_2} - {{\phi}_{F_1}} + {{\phi}_{F_2}}) \right] \nonumber\\
&&{e_{3,3}} =
  2\,\left[ {\sqrt{{G_{F_1}}}}\,\cos ({{\phi}_{F_1}}) -
     {\sqrt{{A_1}\,{G_{F_2}}}}\,\cos ({{\theta }_1} + {{\phi}_{F_2}}) \right] \,
   \left[ {\sqrt{{G_{F_1}}}}\,\cos ({{\phi}_{F_1}}) -
     {\sqrt{{A_2}\,{G_{F_2}}}}\,\cos ({{\theta }_2} + {{\phi}_{F_2}}) \right] \nonumber\\
e_4 = && e_{4,1}\,{T_{{n_{F_1}}}}+{e_{4,2}}\,{T_{{n_{F_2}}}}+
{e_{4,3}}\,{T_{{n_{B_2}}}}\nonumber\\
&& {e_{4,1}} =
   {G_{F_1}}\,\left[ {\sqrt{{G_{F_1}}}} +
     {\sqrt{{A_1}\,{G_{F_2}}}}\,\cos ({{\theta }_1} - {{\phi}_{F_1}} + {{\phi}_{F_2}}) \right] \,
   \left[ {\sqrt{{G_{F_1}}}} + {\sqrt{{A_2}\,{G_{F_2}}}}\,
      \cos ({{\theta }_2} - {{\phi}_{F_1}} + {{\phi}_{F_2}}) \right] \nonumber\\
&&{e_{4,2}} =
 {\sqrt{{A_1}\,{A_2}}}\,{G_{F_2}}\,
   \left[ {\sqrt{{A_1}\,{G_{F_2}}}} + {\sqrt{{G_{F_1}}}}\,
      \cos ({{\theta }_1} - {{\phi}_{F_1}} + {{\phi}_{F_2}}) \right] \,
   \left[ {\sqrt{{A_2}\,{G_{F_2}}}} + {\sqrt{{G_{F_1}}}}\,
      \cos ({{\theta }_2} - {{\phi}_{F_1}} + {{\phi}_{F_2}}) \right] \nonumber\\
&&{e_{4,3}} =
  2\,\left[ {\sqrt{{G_{F_1}}}}\,\cos ({{\phi}_{F_1}}) +
     {\sqrt{{A_1}\,{G_{F_2}}}}\,\cos ({{\theta }_1} + {{\phi}_{F_2}}) \right] \,
   \left[ {\sqrt{{G_{F_1}}}}\,\cos ({{\phi}_{F_1}}) +
     {\sqrt{{A_2}\,{G_{F_2}}}}\,\cos ({{\theta }_2} + {{\phi}_{F_2}}) \right] \nonumber\\
{e_5} =&&
   \frac{1}{2}\,\left( {G_{F_1}} - {A_1}\,{G_{F_2}} \right) \,
    \left( {G_{F_1}} - {A_2}\,{G_{F_2}} \right)\nonumber\\
{e_6} =&&
   {G_{F_1}}\,{T_{{n_{F_1}}}}\,\left[ {G_{F_1}}\,{T_{{n_{F_1}}}} +
4\,{\cos ({{\phi}_{F_1}})}^2\,{T_{{n_{B_2}}}} \right]+  \nonumber\\
+&&{G_{F_2}}\,\sqrt{{A_1}\,{A_2}}{T_{{n_{F_2}}}}\,\left[
\sqrt{{A_1}\,{A_2}}\,{G_{F_2}}\,
        {T_{{n_{F_2}}}} + 4\,{T_{{n_{B_2}}}}\,
        \cos ({{\theta }_1} + {{\phi}_{F_2}})\,
        \cos ({{\theta }_2} + {{\phi}_{F_2}}) +\right.\nonumber\\
+&&\left.2\,{G_{F_1}}\,{T_{{n_{F_1}}}}\,
        \cos ({{\theta }_1} - {{\phi}_{F_1}} + {{\phi}_{F_2}})\,
        \cos ({{\theta }_2} - {{\phi}_{F_1}} + {{\phi}_{F_2}}) \right]
     + 4\,{{T_{{n_{B_2}^2}}}}\nonumber
\end{eqnarray}

\subsubsection{Definitions of $f_j$
coefficients\label{app:fj_coefficients}}

\begin{eqnarray}
f_1 = && 2 \,\tilde I^2\nonumber\\
f_2 = && 2 \,\hat G^2\nonumber\\
f_3 = && f_{3,1}\,T_{n_{F_1}}+f_{3,2}\,T_{n_{F_2}}+f_{3,3}\,T_{n_{B_2}}\nonumber\\
&& {f_{3,1}} =
  \frac{1}{2}\,{G_{F_1}}\,
   {\left[ {\sqrt{{G_{F_1}}}} - {\sqrt{{A_1}\,{G_{F_2}}}}\,
        \cos ({{\theta }_1} - {{\phi}_{F_1}} + {{\phi}_{F_2}}) \right] }^2 \nonumber\\
&&{f_{3,2}} =
   \frac{1}{2}\,{A_1}\,{G_{F_2}}\,
   {\left[ {\sqrt{{A_1}\,{G_{F_2}}}} - {\sqrt{{G_{F_1}}}}\,
        \cos ({{\theta }_1} - {{\phi}_{F_1}} + {{\phi}_{F_2}}) \right] }^2\nonumber\\
&& {f_{3,3}} =
   {\left[ {\sqrt{{G_{F_1}}}}\,\cos ({{\phi}_{F_1}}) -
      {\sqrt{{A_1}\,{G_{F_2}}}}\,\cos ({{\theta }_1} + {{\phi}_{F_2}}) \right] }^2 \nonumber\\
f_4 = && f_{4,1}\,{T_{{n_{F_1}}}}+{f_{4,2}}\,{T_{{n_{F_2}}}}+
{f_{4,3}}\,{T_{{n_{B_2}}}}\nonumber\\
&& {f_{4,1}} =
   \frac{1}{2}\,{G_{F_1}}\,
   {\left[ {\sqrt{{G_{F_1}}}} + {\sqrt{{A_1}\,{G_{F_2}}}}\,
        \cos ({{\theta }_1} - {{\phi}_{F_1}} + {{\phi}_{F_2}}) \right] }^2\nonumber\\
&& {f_{4,2}} =
   \frac{1}{2}\,{A_1}\,{G_{F_2}}\,
   {\left[ {\sqrt{{A_1}\,{G_{F_2}}}} + {\sqrt{{G_{F_1}}}}\,
        \cos ({{\theta }_1} - {{\phi}_{F_1}} + {{\phi}_{F_2}}) \right] }^2\nonumber\\
&&{f_{4,3}} =
  {\left[ {\sqrt{{G_{F_1}}}}\,\cos ({{\phi}_{F_1}}) +
      {\sqrt{{A_1}\,{G_{F_2}}}}\,\cos ({{\theta }_1} + {{\phi}_{F_2}}) \right] }^2\nonumber\\
{f_5} = &&
   \frac{1}{4}\,{\left( {G_{F_1}} - {A_1}\,{G_{F_2}} \right)
   }^2\nonumber\\
{f_6} = &&
   \frac{1}{2}\,{G_{F_1}}\,{T_{{n_{F_1}}}}\,
     \left[ {G_{F_1}}\,{T_{{n_{F_1}}}} +
       4\,{\cos ({{\phi}_{F_1}})}^2\,{T_{{n_{B_2}}}} \right]
       +\nonumber\\
&&+ \frac{1}{2}\,{A_1}\,{G_{F_2}}\,{T_{{n_{F_2}}}}\,
     \left[ 2\,{\cos ({{\theta }_1} - {{\phi}_{F_1}} +
            {{\phi}_{F_2}})}^2\,{G_{F_1}}\,{T_{{n_{F_1}}}} +
       {A_1}\,{G_{F_2}}\,{T_{{n_{F_2}}}} +
       4\,{\cos ({{\theta }_1} + {{\phi}_{F_2}})}^2\,{T_{{n_{B_2}}}}
       \right] + 2\,{{T_{{n_{B_2}}}^2}}\nonumber
\end{eqnarray}

\twocolumn

\section{\label{app:definition_alpha_pars}Definition of
$\alpha$ parameters} The parameters $\alpha_1$ through $\alpha_4$
used in equations (\ref{eq:1st_order_radiometer}),
(\ref{eq:sens_spectrum}), (\ref{eq:kneefreq}) and (\ref{eq:tyeqn})
are defined as:

\begin{eqnarray}
&&{{\alpha }_1} =
   \frac{(1+r)({T_x} - {T_y})}
    {8\left[{T_x} + {T_n} -
      r\,\left( {T_n} + {T_n} \right) \right]}\nonumber\\
&&{{\alpha }_2} =
   \frac{\left( 1 - r \right) \,{T_n}}
    {{T_x} + {T_n} -
      r\,\left( {T_n} + {T_n} \right) }\nonumber\\
&&{{\alpha }_3} =
   {T_n}\,\left( \frac{1}
       {{T_x} + {T_n}} +
      \frac{1}{{T_y} + {T_n}} \right)\nonumber\\
&&{{\alpha }_4} =
   \frac{T_x-T_y}
       {{T_y} + {T_n}}
\label{eq:alpha_1_to_4}
\end{eqnarray}

\end{document}